# Two-plasmon decay instability and 3/2 harmonic generation in inhomogeneous magnetized plasma

S. S. Ghaffari-Oskooei[1], A. A. Molavi Choobini[2*]

[1]Department of Atomic and Molecular Physics, Faculty of Physics, Alzahra University, Tehran, Iran,

[2] Department of Physics, University of Tehran, Tehran, 14399-55961, Iran.

**Abstract:**

Two-plasmon decay instability emerges as the parametric decay of laser beams into two plasma waves which is expected for hohlraum in inertial confinement fusion. The behavior of this instability in magnetized plasma is investigated in the present study where the thermal effects are considered. The study demonstrates that the applying of a dc magnetic field significantly increases the rate at which the instability develops, while the increase of electron temperature has the opposite effect. The generation of 3⁄2 harmonic, which results from the beating of the incident laser and plasma wave, is also studied in the present study. The intensity of this harmonic, which appears in the spectrum as a sideband derived analytically, indicates that an increase in the dc magnetic field increases the intensity of this sideband.

*Corresponding author E-mail address: aa.molavich@ut.ac.ir

I. **Introduction**

The propagation of laser pulses within a plasma medium is a critical area of study in plasma physics, given its significance across various applications, such as THz wave emission, inertial confinement fusion (ICF), harmonic generation, particle acceleration, and X-ray production [1-5]. These applications underscore the necessity of understanding laser-plasma interactions and the associated phenomena. The interaction between lasers and plasmas gives rise to a range of nonlinear phenomena collectively referred to as parametric instabilities. Examples of these instabilities include Raman scattering, two-plasmon decay, and Brillouin scattering, commonly observed in laser-plasma interactions [6-8]. The ability to control the growth rate of these instabilities is crucial for ensuring the stable propagation of laser pulses in plasma environments. Both theoretical and experimental investigations have been extensively conducted to analyze these parametric instabilities, which are particularly relevant in the context of the ICF experiments [9-10]. In two-plasmon decay (TPD), a high-energy electromagnetic wave, called the pump wave, breaks down into two lower-energy plasma waves, known as daughter waves [11]. According to the principles of momentum and energy conservation, TPD occurs when the frequency of the incident electromagnetic wave is twice that of the plasma wave. This condition suggests that TPD is likely to occur when the plasma density is approximately one-quarter of the critical density. In the context of ICF experiments, this instability is particularly significant as it leads to the generation of hot electrons and the preheating of the target. As a result, a thorough analysis of the corresponding unstable modes, including both absolute and convective modes, is essential [12].

The behavior of absolute and convective modes of TPD in plasmas has been the subject of numerous studies. Liu and Rosenbluth derived an analytical formula for the growth rate of TPD in an inhomogeneous plasma, considering a linear density profile in their analysis [13]. Xin-Feng and co-workers examined the absolute and convective modes of TPD in magnetized plasmas, focusing on the interaction between an incident electromagnetic wave in the extraordinary mode and plasma waves [14]. Ji and the team investigated the occurrence of TPD resulting from the interaction between a twisted laser beam and plasma waves [15]. Bedros et al. proposed a unified theory that encompasses both stimulated Raman scattering and TPD in inhomogeneous plasmas [16]. DuBois and co-workers conducted a numerical investigation of the polar pattern, density-averaged angular distribution, and power spectrum of the secondary radiation generated by TPD [17]. Their findings are consistent with experimental observations reported in Ref. [18]. Additionally, Liao and colleagues documented the generation of terahertz (THz) waves during laser-solid interactions, attributing this radiation to the effects of TPD [19]. E. Z. Gusakov and A. Yu. Popov investigated a method to reduce the saturation level of the plasmon parametric decay instability in extraordinary mode waves [20]. They indicated that the confinement of a high harmonic ion Bernstein (IB) wave inside a magnetic island results in a diminished saturation amplitude of the primary instability. E. Rovere and colleagues investigated the scaling of hot electrons in CF plasmas for two-plasmon decay [21]. They developed a scaling law that predicts the generation of hot electrons through the two-plasmon decay process at a plasma density equal to one-fourth of the critical density. This model considers the effects of Langmuir decay instabilities and the depletion of the laser pump wave. A study of parametric decay instabilities in plasmas using kinetic theory, focusing on the region near the upper hybrid resonance has been studied by Jiangyue Han and co-workers [22]. In a uniform plasma, they numerically solved the kinetic nonlinear dispersion relation for parametric decay instabilities using parameters that correspond to electron cyclotron heating experiments performed on the ASDEX-U tokamak.

This study presents a groundbreaking investigation into Two-Plasmon Decay (TPD) within a magnetized, inhomogeneous plasma. A unique configuration is explored where the applied magnetic field aligns parallel to the incident laser wave vector. The analysis incorporates the critical influence of electron temperature, providing a more comprehensive understanding of TPD dynamics in this complex environment. Novel analytical expressions for TPD growth rate and convective gain are derived under a linear density profile approximation. Moreover, the study investigates the intensity of the 3/2 harmonic radiation generated as a result of TPD in the plasma and provides important implications for experimental and applied plasma physics. The present article is organized as follows: In section II, the growth rate and convective gain of two-plasmon decay in inhomogeneous plasma with linear density profile is derived analytically through fluid theory of plasmas. The intensity of 3/2 harmonic radiation, which is generated through two-plasmon decay, is also investigated analytically in this section. Section III is devoted to the numerical examination of this parametric instability in inhomogeneous plasma where the effect of laser and plasma parameters is investigated. Conclusions are drawn in section IV.

## II. Theoretical Model

A circularly polarized laser beam with the frequency and wavenumber of $\omega_0$ and $k_0$ that is incident on a magnetized inhomogeneous plasma slab is considered. The density profile $n_0(z) = n_0(1 + \frac{z}{L})$ is considered here. The plasma slab is immersed in a constant magnetic field of $\vec{B}_0 = B_0 \hat{e}_z$. According to the fluid theory of plasmas, electron density and its velocity satisfy the following equations:

$$m\frac{d\vec{v}}{dt} = -e\vec{E} - \frac{e}{c}\vec{v} \times (\vec{B}_0 + \vec{B}) - \frac{3v_{th}^2 \vec{\nabla} N}{N} \tag{1}$$

$$\frac{\partial N}{\partial t} + \vec{\nabla}.(N\vec{v}) = 0 \tag{2}$$

where $\vec{B}$ and $\vec{E}$ represent the magnetic and electric fields. The parameters of $e$, $N$, $\vec{v}$, $m$, and $v_{th}$ are the charge, density, velocity, mass and thermal velocity of the electrons, respectively. The electric field of the incident laser beam is given by:

$$\vec{E}_L = \frac{1}{2}(\hat{e}_x + i\sigma\hat{e}_y)E_0 e^{-i\omega_0 t + ik_0 z} + c.c. \tag{3}$$

where $E_0$ denotes the magnitude of the electric field associated with the laser. The parameter of $\sigma = \pm 1$ indicates the right-hand or left-hand polarization of the laser. By considering $N = n_0 + n$ and $\vec{v} = \vec{v}_0 + \vec{v}_1$, the following equations are derived:

$$\vec{v}_0 = -i\frac{e}{m(\omega_0 - \sigma\omega_c)}\vec{E}_L \tag{4}$$

$$\frac{\partial \vec{v}_1}{\partial t} = -\frac{e}{m}\vec{E} - 3v_{th}^2 \frac{\vec{\nabla} n}{n_0} + i\sigma\omega_c \vec{v}_1 - \vec{\nabla}(\vec{v}_0.\vec{v}_1) \tag{5}$$

$$\frac{\partial n}{\partial t} + \vec{\nabla}.(n_0 \vec{v}_1) + \vec{v}_0.\vec{\nabla} n = 0 \tag{6}$$

By setting $\vec{v}_1 = \vec{\nabla}\psi$ and $\vec{E} = -\vec{\nabla}\phi$, the quantities $\phi$, $n$ and $\psi$ are defined for the analysis of two-plasmon decay instability in the magnetized plasma as follows:

$\phi(x,y,z,t) =$

$\frac{1}{2}\phi_1(z,t)e^{i\omega_1 t - ik_x x - ik_y y - i\int_0^z k_{z1}(z')dz'} + \frac{1}{2}\phi_2(z,t)e^{i\omega_2 t - ik_x x - ik_y y - i\int_0^z k_{z2}(z')dz'} + c.c.$ (7)

$n(x,y,z,t) =$

$\frac{1}{2}n_1(z,t)e^{i\omega_1 t - ik_x x - ik_y y - i\int_0^z k_{z1}(z')dz'} + \frac{1}{2}n_2(z,t)e^{i\omega_2 t - ik_x x - ik_y y - i\int_0^z k_{z2}(z')dz'} + c.c.$ (8)

$\psi(x,y,z,t) =$

$$\frac{1}{2}\psi_1(z,t)e^{i\omega_1 t - ik_x x - ik_y y - i\int_0^z k_{z1}(z')dz'} + \frac{1}{2}\psi_2(z,t)e^{i\omega_2 t - ik_x x - ik_y y - i\int_0^z k_{z2}(z')dz'} + c.c. \quad (9)$$

Insertion of Eqs. (7)-(9) in Eqs. (4)-(6) along with Poisson's equation, $\nabla^2\phi = 4\pi en$, leads to the following equations:

$$\frac{\partial}{\partial t}\psi_1 + i(\omega_1 - \sigma\omega_c)\psi_1 - \frac{i(k_x+k_y)v_0}{2}\psi_2^* e^{-i\delta} = \frac{e}{m}\phi_1 - 3v_{th}^2\frac{n_1}{n_0} \quad (10)$$

$$\frac{\partial}{\partial t}\psi_2 + i(\omega_2 - \sigma\omega_c)\psi_2 + \frac{i(k_x+k_y)v_0}{2}\psi_1^* e^{-i\delta} = \frac{e}{m}\phi_2 - 3v_{th}^2\frac{n_2}{n_0} \quad (11)$$

$$\frac{\partial}{\partial t}n_1 + i\omega_1 n_1 - \frac{i(k_x+k_y)v_0}{2}n_2^* e^{-i\delta} = n_0\left[k_1^2\psi_1 + i\frac{\partial k_{z1}}{\partial z}\psi_1\right] - \frac{\partial n_0}{\partial z}\left(\frac{\partial \psi_1}{\partial z} - ik_{z1}\psi_1\right) \quad (12)$$

$$\frac{\partial}{\partial t}n_2 + i\omega_2 n_2 + \frac{i(k_x+k_y)v_0}{2}n_1^* e^{-i\delta} = n_0\left[k_2^2\psi_2 + i\frac{\partial k_{z2}}{\partial z}\psi_2\right] - \frac{\partial n_0}{\partial z}\left(\frac{\partial \psi_2}{\partial z} - ik_{z2}\psi_2\right) \quad (13)$$

$$\left(k_1^2 + i\frac{\partial k_{z1}}{\partial z}\right)\phi_1 + 2ik_{z1}\frac{\partial \phi_1}{\partial z} = -4\pi e n_1 \quad (14)$$

$$\left(k_2^2 + i\frac{\partial k_{z2}}{\partial z}\right)\phi_2 + 2ik_{z2}\frac{\partial \phi_2}{\partial z} = -4\pi e n_2 \quad (15)$$

where

$$\delta = \int_0^z (k_0 - k_{z1} - k_{z2})dz' \quad (16a)$$

$$k_1^2 = k_x^2 + k_y^2 + k_{z1}^2 \quad (16b)$$

$$k_2^2 = k_x^2 + k_y^2 + k_{z2}^2 \quad (16c)$$

Using Eqs. (14) and (15), the elimination of the variables $n_1$ and $n_2$ from the system of equations represented by Eqs. (10) and (11), one yielding as:

$$\frac{\partial}{\partial t}\psi_1 + i(\omega_1 - \sigma\omega_c)\psi_1 - \frac{i(k_x+k_y)v_0}{2}\psi_2^* e^{-i\delta} = \frac{e}{m}\phi_1 + \frac{3v_{th}^2}{4\pi en_0}\left(k_1^2 + \frac{i\partial k_1}{\partial z} + 2ik_1\frac{\partial}{\partial z}\right)\phi_1 \quad (17a)$$

$$\frac{\partial}{\partial t}\psi_2 + i(\omega_2 - \sigma\omega_c)\psi_2 - \frac{i(k_x+k_y)v_0}{2}\psi_1^* e^{-i\delta} = \frac{e}{m}\phi_2 + \frac{3v_{th}^2}{4\pi en_0}\left(k_2^2 + \frac{i\partial k_2}{\partial z} + 2ik_2\frac{\partial}{\partial z}\right)\phi_2 \quad (17b)$$

Using zeroth-order solution of $\psi_1$ and $\psi_2$ and merging Eqs. (17a)-(17b) lead to the following equations:

$$\frac{\partial \phi_1}{\partial t} + \frac{3k_{z1}v_{th}^2}{(\omega_1 - \sigma\omega_c)}\frac{\partial \phi_1}{\partial z} - \left[\frac{(\omega_1 - \sigma\omega_c)}{2n_0 k_1^2}\frac{\partial n_0}{\partial z}k_{z1} + \frac{1}{4k_{z1}n_0(\omega_1 - \sigma\omega_c)}\frac{\partial n_0}{\partial z}\right]\phi_1$$

$$= -\frac{1}{4}i(k_x + k_y)v_0\left[\frac{(\omega_2-\sigma\omega_c)}{(\omega_1-\sigma\omega_c)} - \frac{k_2^2}{k_1^2}\right]\phi_2^*\exp(-i\delta) \quad (18a)$$

$$\frac{\partial \phi_2}{\partial t} + \frac{3k_{z1}v_{th}^2}{(\omega_2 - \sigma\omega_c)}\frac{\partial \phi_2}{\partial z} - \left[\frac{(\omega_2 - \sigma\omega_c)}{2n_0 k_2^2}\frac{\partial n_0}{\partial z}k_{z2} + \frac{1}{4k_{z2}n_0(\omega_2 - \sigma\omega_c)}\frac{\partial n_0}{\partial z}\right]\phi_2$$

$$= \frac{1}{4}i(k_x + k_y)v_0\left[\frac{(\omega_1-\sigma\omega_c)}{(\omega_2-\sigma\omega_c)} - \frac{k_1^2}{k_2^2}\right]\phi_1^*\exp(-i\delta) \quad (18b)$$

By defining

$$a_1 = \sqrt{\frac{|v_0|(k_x+k_y)}{4}\left|\frac{\omega_1-\sigma\omega_c}{\omega_2-\sigma\omega_c} - \frac{k_1^2}{k_2^2}\right|}\phi_1^* \tag{19a}$$

$$a_2 = \sqrt{\frac{|v_0|(k_x+k_y)}{4}\left|\frac{\omega_2-\sigma\omega_c}{\omega_1-\sigma\omega_c} - \frac{k_2^2}{k_1^2}\right|}\phi_2^* \tag{19b}$$

and phenomenologically adding of damping rates, one can derive the following equations:

$$\left(\frac{\partial}{\partial t} + v_1 + \gamma_1\right)a_1 = \Gamma a_2^* \exp(-i\delta) \tag{20a}$$

$$\left(\frac{\partial}{\partial t} + v_2 + \gamma_2\right)a_2 = \Gamma a_1^* \exp(-i\delta) \tag{20b}$$

where

$$\Gamma^2 = \frac{v_0^2(k_x+k_y)^2}{16}\left[\frac{\omega_2-\sigma\omega_c}{\omega_1-\sigma\omega_c} - \frac{k_2^2}{k_1^2}\right]\left[-\frac{\omega_1-\sigma\omega_c}{\omega_2-\sigma\omega_c} + \frac{k_1^2}{k_2^2}\right] \tag{21a}$$

$$\gamma_1 = v_1 - \frac{(\omega_1-\sigma\omega_c)}{2n_0 k_1^2}\frac{\partial n_0}{\partial z}k_{z1} - \frac{1}{4k_{z1}n_0(\omega_1-\sigma\omega_c)}\frac{\partial n_0}{\partial z} \tag{21b}$$

$$\gamma_2 = v_2 - \frac{(\omega_2-\sigma\omega_c)}{2n_0 k_2^2}\frac{\partial n_0}{\partial z}k_{z2} - \frac{1}{4k_{z2}n_0(\omega_2-\sigma\omega_c)}\frac{\partial n_0}{\partial z} \tag{21c}$$

The convective gain of two-plasmon decay is defined as follows:

$$g_R = \exp(\pi\Lambda) \tag{22}$$

where

$$\Lambda = \frac{|\Gamma|^2}{v_1 v_2 \left(\frac{d\kappa}{dz}\right)} \tag{23}$$

the parameter of $\kappa(z)$ is the mismatch of three wavevectors:

$$\kappa(z) = k_0 - k_{z1} - k_{z2} \tag{24}$$

and derivative of $\kappa(z)$ is given by:

$$\frac{d\kappa(z)}{dz} = \left(-\frac{1}{2k_0 c^2\left(1-\frac{\sigma\omega_c}{\omega_0}\right)} + \frac{1}{6v_{th}^2 k_{z1}} + \frac{1}{6v_{th}^2 k_{z2}}\right)\frac{4\pi e^2}{m}\frac{\partial n_0}{\partial z} \tag{25}$$

Therefore, the convective gain is expressed as:

$$\Lambda = \frac{mv_0^2(k_x+k_y)^2}{64\pi e^2 v_1 v_2 \frac{\partial n_0}{\partial z}}\left[\frac{\omega_2-\sigma\omega_c}{\omega_1-\sigma\omega_c} - \frac{k_2^2}{k_1^2}\right]\left[-\frac{\omega_1-\sigma\omega_c}{\omega_2-\sigma\omega_c} + \frac{k_1^2}{k_2^2}\right]\left(-\frac{1}{2k_0 c^2\left(1-\frac{\sigma\omega_c}{\omega_0}\right)} + \frac{1}{6v_{th}^2 k_{z1}} + \frac{1}{6v_{th}^2 k_{z2}}\right)^{-1} \tag{26}$$

By setting $\omega_c = 0$, Eqs. (21a) and (26) can be reduced to match the findings presented in Ref. [12]. The beating of plasma wave and incident laser leads to the generation of a scattered wave as described follows:

$$\vec{E}_s = \frac{1}{2} E_s e^{-i\vec{k}_s \cdot \vec{r} + i\omega_s t} + c.c. \tag{27}$$

where $\omega_s$ and $\vec{k}_s$ denote the frequency and wavevector of the scattered waves, respectively. Their expressions are as follows:

$$\omega_s = \omega_1 + \omega_0 \tag{28}$$

$$\vec{k}_s = \vec{k}_0 + \vec{k}_1 \tag{29}$$

The frequency of the scattered wave is $\omega_s = \frac{3}{2}\omega_0$ under the conditions $\omega_1 = \omega_p$ and $\omega_0 = 2\omega_p$. Generation of the $\frac{3}{2}\omega_0$ radiation due to two-plasmon decay instability is a notable consequence observed in experimental studies [18]. The intensity of the scattered waves can be derived analytically through Maxwell's wave equations:

$$-\nabla^2 \vec{E}_s + \frac{1}{c^2} \frac{\partial^2 \vec{E}_s}{\partial t^2} = -\frac{4\pi}{c^2} \vec{J} \tag{30}$$

where $\vec{J}$ is the current density generated by ponderomotive forces as:

$$[c^2 \nabla^2 - \frac{\partial^2}{\partial t^2} - \frac{\omega_p^2}{\left(1 - \frac{\sigma \omega_c}{\omega_s}\right)}] E_s = \frac{k\omega_p^2}{4\pi e n_0 (1 - \frac{\sigma \omega_c}{\omega_0})} \vec{E}_0 \cdot \vec{E}_1 \tag{31}$$

By using dispersion relation, $c^2 k_s^2 = \omega_s^2 - \frac{\omega_p^2}{(1 - \frac{\sigma \omega_c}{\omega_s})}$, and applying the derivatives in Eq. (31), one can derive the following equation:

$$2ic^2 k_{sz} \frac{\partial E_s}{\partial z} = \frac{\delta n}{\sqrt{2} n_0} E_0 \omega_p^2 e^{i\kappa_1' z^2} \tag{32}$$

where the mismatch is determined by the expression:

$$\kappa_1' = \frac{2\pi e^2}{3 m v_{th}^2 k_{z1}} \frac{\partial n_0}{\partial z} \tag{33}$$

Therefore, the intensity of scattered waves is given by:

$$I_s = I_0 \left(\frac{\omega_p^2}{2\sqrt{2} k_{sz} \kappa_1' c^2}\right)^2 \left(\frac{\delta n}{n_0}\right)^2 \tag{34}$$

where $I_0$ is the intensity of the incident laser.

### III. Results and Discussion

Investigation of instabilities is crucial for advancing knowledge across various plasma-related fields. In inertial confinement fusion (ICF) targets, the matching condition is met near the critical-density surface, where energetic electrons generated by these instabilities can preheat the target. This preheating disrupts the low-adiabatic conditions required for ICF compression, thereby degrading the implosion. The two-plasmon decay (TPD) instability is particularly worrisome in direct-drive inertial confinement fusion (ICF) because it can be easily triggered (low threshold) and produces electrons with high kinetic energy, which can degrade the performance of the ICF target. Beyond ICF, laser-plasma instabilities also play a significant role in high-energy-density physics, THz wave emission, particle acceleration, and the development of advanced light sources. Effective control of these instabilities can enhance the performance of laser-driven particle accelerators. Moreover, intense laser interactions with plasmas can produce extreme conditions that are valuable for studying high-pressure materials, potentially leading to breakthroughs in understanding material properties under conditions similar to those found in planetary interiors. The simulations of two-plasmon decay instability and 3/2 harmonic generation in inhomogeneous magnetized plasma are investigated.

Figures 1, 2, and 3 demonstrate the variations of the rate of increase of the two-plasmon decay instability, the convective gain ($g_R$) (or Rosenbluth gain), and Λ of two-plasmon decay versus the normalized transverse wave number ($k_\perp/k_0$) for different cyclotron frequencies, respectively. They show that with the increase of cyclotron frequency, these parameters enhance. The shape of the changes for the growth rate of two-plasmon decay and Λ increases equally until they reach saturation. As the cyclotron frequency increases, the magnetic field more strongly influences electron motion in the plasma. This influence enhances the coupling between the laser-driven electric field and the plasma waves, which is the key mechanism behind the two-plasmon decay (TPD) instability. As this frequency increases, the electrons become more tightly confined to their gyro-orbits. This confinement reduces the ability of the electrons to escape the interaction region, which in turn increases the efficiency of energy transfer from the laser field to the plasma waves, leading to an enhanced growth rate. In addition, for TPD, the growth rate is sensitive to resonance conditions between the laser field and the plasma waves. As the magnetic field increases, the conditions for resonance improve, leading to a stronger, more rapid amplification of the instability until saturation occurs. Saturation occurs when the energy transfer to the plasma waves reaches a balance with dissipative processes or when nonlinear effects become significant, limiting further growth of the instability. This saturation is typically due to the depletion of available energy in the system or the onset of wave-breaking or other nonlinear effects that prevent further exponential growth. However, around the normalized wave number $k_\perp/k_0 \approx 0.65$ the coupling between the Langmuir waves, which drive the two-plasmon decay, and the magnetic field changes. The trend is reversed for Λ, and so, the magnitude of it decreases with increasing cyclotron frequency. The growth rate describes how quickly the instability amplifies locally, while the Λ involves both the growth rate and the spatial extent of the instability, i.e., how far the wave travels before dissipating. These two quantities are related but are influenced differently by the plasma conditions, especially the magnetic field strength. As the cyclotron frequency increases, the influence of the magnetic field on the electron motion becomes more pronounced. This can enhance the growth of the TPD instability initially, as seen in the growth rate. However, when it comes to the Λ, propagation of the instability may be restricted by the stronger magnetic confinement. This would reduce the Λ despite the increased growth rate.

The impact of intensity of the incident laser on variations of rate of growth of the two-plasmon decay instability and the Λ versus the normalized transverse wave number ($k_\perp/k_0$) are depicted in Figures 3 and 4, respectively. The figures indicated that as the intensity of the parameter enhances, the growth rate of two-plasmon decay increases. The intensity of the laser is proportional to the square of its electric field. As the intensity increases, the strength of the electric field interacting with the plasma also increases. This stronger electric field influences the resonance conditions required for the TPD instability and amplifies the forces acting on the electrons in the plasma. As a result, this enhances the energy transfer between the laser and the plasma waves, further enhancing the growth rate of the instability. Furthermore, the TPD instability has a threshold for excitation, which is dependent on the intensity of the incident laser. As the laser intensity rises, the nonlinearity of the plasma response becomes more pronounced, driving the growth of the Langmuir waves more effectively. This increased nonlinearity leads to a faster development of the TPD instability. While again, around $k_\perp/k_0 \approx 0.65$, the trend of $\Lambda$ reverses and its magnitude decreases with increasing intensity. The reversal in the trend at higher intensities is driven by the onset of nonlinear effects, such as wave-breaking, which inhibit the propagation of the plasma waves, even as the local growth rate continues to increase due to stronger laser-plasma coupling. These processes can redistribute energy within the plasma, leading to less efficient amplification of the instability over large distances. As a result, while the local growth rate (which reflects the initial amplification) continues to rise, the $\Lambda$ (which accounts for the spatial extent and net amplification of the instability) decreases. Therefore, these nonlinear effects reduce the efficiency of wave propagation, causing the $\Lambda$ to decrease despite the continued increase in growth rate.

Figures 6 and 7 display the role of polarization states on variations of rate of growth of the two-plasmon decay instability and the $\Lambda$ versus the normalized transverse wave number $(k_\perp/k_0)$, individually. As the figures show, for a laser beam with a right-handed circular polarization state, as opposed to left-handed circular polarization or linear polarization, the magnitude of these parameters is the largest. The reason why right-handed circular polarization (RHCP) exhibits a larger two-plasmon decay (TPD) growth rate and $\Lambda$ compared to left-handed circular polarization (LHCP) and linear polarization is tied to the interaction between the polarization of the laser field and the plasma's response, particularly in the context of magnetized plasmas. In magnetized plasma, the interaction between the laser field and the plasma depends heavily on the polarization of the incident laser. Right-handed circularly polarized light (RHCP) interacts more effectively with electrons in a plasma in the presence of an external magnetic field when the polarization matches the electron's natural gyration direction. In this situation, the electrons experience a resonant interaction with the laser electric field, leading to more efficient energy absorption. This enhanced interaction increases the energy available to drive the TPD instability, resulting in a higher growth rate and $\Lambda$. In other words, in magnetized plasma, the magnetic field introduces a preferred handedness for wave-particle interactions. For RHCP, the electric field of the laser rotates in the same direction as the electron's cyclotron motion. This alignment leads to more effective energy transfer from the laser to the plasma, enhancing the growth of the TPD instability. In contrast, LHCP (where the polarization rotates opposite to the electron's cyclotron motion) and linear polarization do not couple as efficiently with the plasma. In LHCP, the electric field and the electron's motion are out of phase, leading to less resonant energy transfer. For linear polarization, the electric field does not rotate, resulting in weaker interactions with the gyrating electrons. Furthermore, the nonlinear nature of the TPD instability means that small differences in energy coupling can result in significant differences in the growth rate and $\Lambda$. Because RHCP couples more efficiently with the plasma, it can more rapidly drive the growth of Langmuir waves. The efficiency of exciting the Langmuir waves responsible for TPD depends on the symmetry of the electric field in the laser-plasma interaction. The circular symmetry of RHCP can more effectively drive coherent wave excitations in the plasma, leading to stronger instability. Meanwhile, linear polarization, with its alternating field direction, can be less effective in driving coherent plasma waves. Considering that the reversal process in the trend of $\Lambda$ around $k_\perp/k_0 \approx 0.65$ also is in various polarization states due to the mentioned reasons above.

The variations in rate of growth of the two-plasmon decay instability and $\Lambda$ versus the normalized transverse wave number ($k_\perp/k_0$), are investigated and depicted in Figs. 8 and 9 for different temperatures. As the figures show, with the increase of thermal energy, the magnitude of parameters the growth rate of two-plasmon decay, and the $\Lambda$ of two-plasmon decay decreases. Also, the width of the graph for the parameter of the growth rate of two-plasmon decay is reduced. For the two-plasmon decay instability to grow, a strong and coherent interaction between the Langmuir waves and the electrons is necessary. The decrease in the growth rate of two-plasmon decay and $\Lambda$ with increasing thermal energy is due to the disruptive effects of thermal motion, which weakens the coherence of wave-particle interactions. As the thermal energy of the plasma increases, the random motion of the electrons increases as well. This increased thermal motion causes the electron velocity distribution to broaden, as the result the phase matching between the electrons and Langmuir waves over a wider range of velocities distributed. This causes in turn weakens the coherence between the plasma waves and the electron population responsible for driving the two-plasmon decay instability. This leads to a reduced energy transfer from the electron population to the Langmuir waves, causing a decrease in the TPD growth rate and $\Lambda$.

Figure 10 illustrates the variations of the normalized intensity of scattered waves versus the normalized frequency for magnetized and unmagnetized states and various plasma densities. According to the figure, the scattered intensity in the magnetized state is greater than in the non-magnetized state. Also, with the increase in the density of plasma electrons in the magnetized state, the magnitude of the normalized intensity increases. In applying of a dc magnetic field, the plasma responds differently to electromagnetic waves compared to an unmagnetized state. The magnetic field creates conditions that enhance the coupling between the incident waves and the plasma, leading to stronger excitation of plasma waves (such as Langmuir waves and electron cyclotron waves). This enhanced coupling increases the energy transferred from the incident laser to the plasma, resulting in higher scattered intensities in the magnetized state. Furthermore, in a magnetized plasma, electrons experience cyclotron motion due to the magnetic field. When the frequency of the incident wave or its harmonics resonates with the electron cyclotron frequency, the energy transfer between the wave and the electrons becomes much more efficient. This resonant interaction boosts the scattered intensity because more energy is absorbed from the wave and subsequently re-emitted as scattered waves. As the plasma electron density increases, the number of particles available for wave interactions increases as well. This leads to stronger scattering because there are more electrons to participate in the interaction with the incident wave, resulting in a larger scattered intensity. In denser plasmas, collective effects (where large numbers of electrons oscillate together) become more pronounced. These collective effects enhance the scattering process, especially in applying of a dc magnetic field, leading to higher scattered intensities as the density increases. In magnetized plasmas, these effects are further amplified by the magnetic field, resulting in higher scattered intensities compared to the unmagnetized state.

## IV. Conclusions

Two-plasmon decay in magnetized plasma is investigated in the present study by an analytic approach based on the fluid theory of plasmas. Analytical expressions are derived for growth rate and $\Lambda$ of TPD where the effect of dc magnetic field and plasma temperature is considered. The effect of laser intensity and its polarization, as well as cyclotron frequency and electron temperature on growth rate and $\Lambda$ of two-plasmon decay, is examined numerically. Curves of the growth rate of TPD versus the transverse wavevector for different laser and plasma parameters obey a similar trend. An increase in the dc magnetic field enhances the rate of growth of the two-plasmon decay instability indicating that applying of a dc magnetic field, which can be provided by external magnets or self-generated magnetic fields generated by intense laser pulses, plays a crucial role in this instability. An increase of laser intensity enhances the growth rate of TPD which indicates that intense laser pulses are more susceptible to this instability. On the other hand, the increase in electron temperature attenuates the growth rate of instability. Convective growth rate or Rosenbluth gain ($g_R$) and the parameter $\Lambda$ are also examined numerically where the effect of laser intensity, electron temperature, cyclotron frequency, and polarization states are considered. An increase in the dc magnetic field amplifies the convective growth rate ($g_R$) while its effect on parameter $\Lambda$ is not a continuous amplification. There is a point in the curve of $\Lambda$ where the amplification trend reverses which means that the increase of dc magnetic field decreases $\Lambda$. A similar behavior is also observed in the increase in laser intensity. The generation of $3/2$ harmonic, which is a consequence of TPD, is studied analytically in the present study. An increase in dc magnetic and plasma density enhances the intensity of this harmonic.

**Acknowledgment**

This research did not receive any specific grant from funding agencies in the public, commercial, or not-for-profit sectors.

## List of Figures & Captions

**Fig. 1.** Variations of the growth rate of two-plasmon decay versus the normalized transverse wave number $(k_\perp/k_0)$ for different cyclotron frequencies, $I = 2 \times 10^{14}\, W/cm^2$, $\sigma = +1$, $k_B T_e = 1 keV$ and $n_p = 2 \times 10^{16}\, m^{-3}$.

**Fig. 2.** The impact of various cyclotron frequencies on variations of convective gain $(g_R)$ (or Rosenbluth gain) versus the normalized transverse wave number $(k_\perp/k_0)$ for $I = 2 \times 10^{14}\, W/cm^2$, $\sigma = +1$, $k_B T_e = 1 keV$ and $n_p = 2 \times 10^{16}\, m^{-3}$.

**Fig. 3.** The effect of various cyclotron frequencies on variations of $\Lambda$ versus the normalized transverse wave number $(k_\perp/k_0)$ for $I = 2 \times 10^{14}\, W/cm^2$, $\sigma = +1$, $k_B T_e = 1 keV$ and $n_p = 2 \times 10^{16}\, m^{-3}$.

**Fig. 4.** Variations of the growth rate of two-plasmon decay versus the normalized transverse wave number $(k_\perp/k_0)$ for different intensities of the incident laser, $\omega_c/\omega_0 = 0.01$, $\sigma = +1$, $k_B T_e = 1 keV$ and $n_p = 2 \times 10^{16}\, m^{-3}$.

**Fig. 5.** The effect of various intensities of the incident laser on variations of $\Lambda$ versus the normalized transverse wave number $(k_\perp/k_0)$ for $\omega_c/\omega_0 = 0.01$, $\sigma = +1$, $k_B T_e = 1 keV$ and $n_p = 2 \times 10^{16}\, m^{-3}$.

**Fig. 6.** Variations of the growth rate of two-plasmon decay versus the normalized transverse wave number $(k_\perp/k_0)$ for different polarization states, $\omega_c/\omega_0 = 0.01$, $I = 2 \times 10^{14}\, W/cm^2$, $k_B T_e = 1 keV$ and $n_p = 2 \times 10^{16}\, m^{-3}$.

**Fig. 7.** The role of various polarization states on variations of $\Lambda$ versus the normalized transverse wave number $(k_\perp/k_0)$ for $\omega_c/\omega_0 = 0.01$, $I = 2 \times 10^{14}\, W/cm^2$, $k_B T_e = 1 keV$ and $n_p = 2 \times 10^{16}\, m^{-3}$.

**Fig. 8.** Variations of the growth rate of two-plasmon decay versus the normalized transverse wave number $(k_\perp/k_0)$ for different thermal energy, $\omega_c/\omega_0 = 0.01$, $I = 2 \times 10^{14}\, W/cm^2$, $\sigma = +1$ and $n_p = 2 \times 10^{16}\, m^{-3}$.

**Fig. 9.** The impact of various temperatures on variations of $\Lambda$ versus the normalized transverse wave number $(k_\perp/k_0)$ for $\omega_c/\omega_0 = 0.01$, $I = 2 \times 10^{14}\, W/cm^2$, $\sigma = +1$ and $n_p = 2 \times 10^{16}\, m^{-3}$.

**Fig. 10.** Variations of the normalized intensity of scattered waves versus the normalized frequency for magnetized and unmagnetized states and various plasma densities, $I = 2 \times 10^{14}\, W/cm^2$, $\sigma = +1$ and $k_B T_e = 1 keV$.

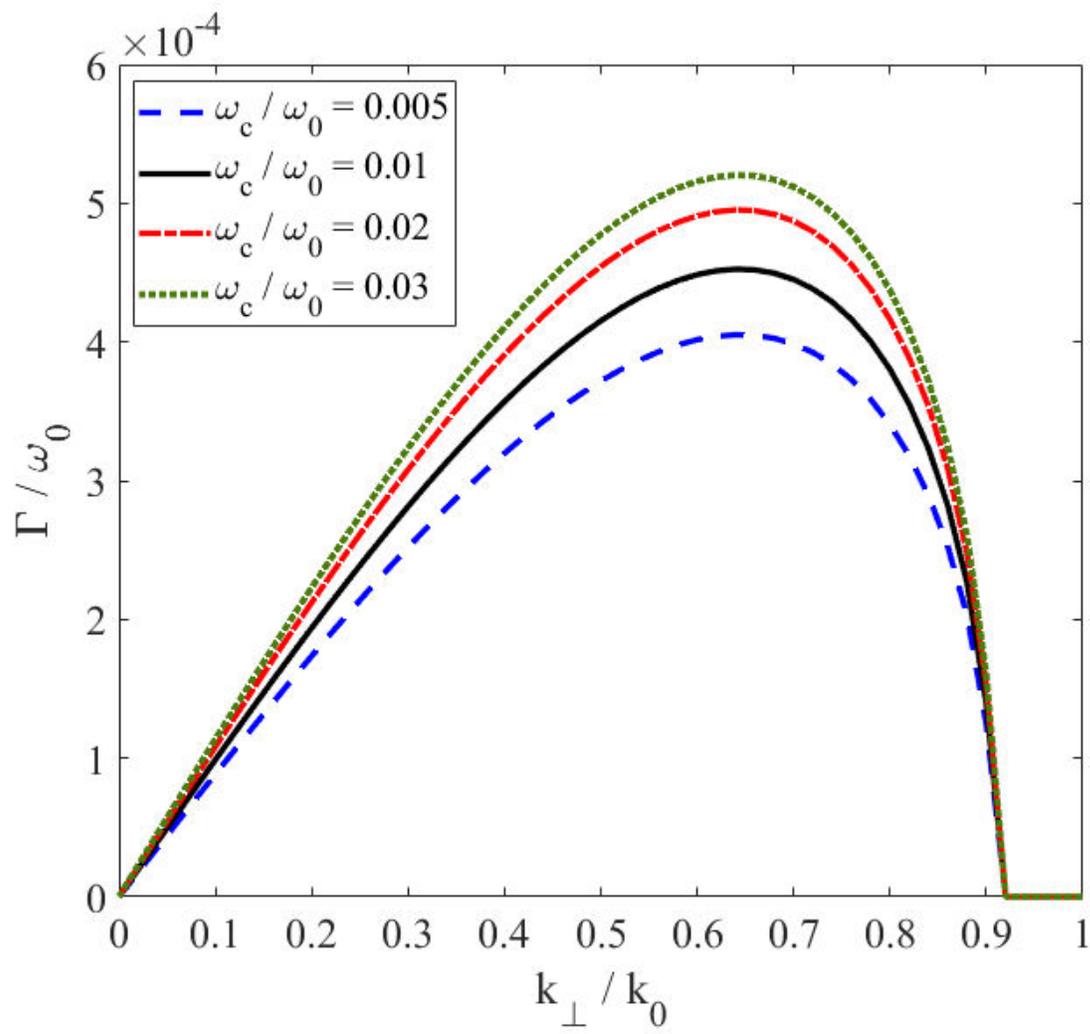

**Fig. 1**

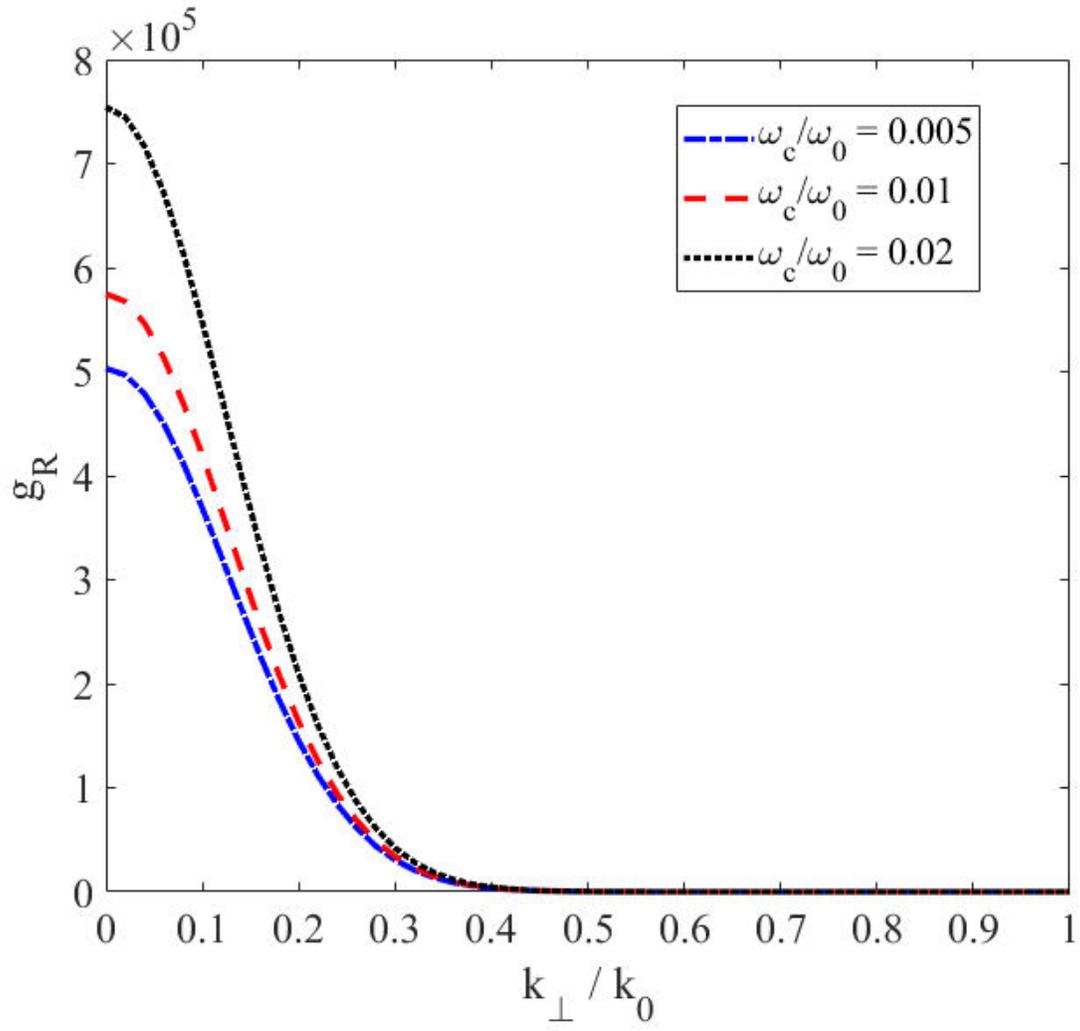

**Fig. 2**

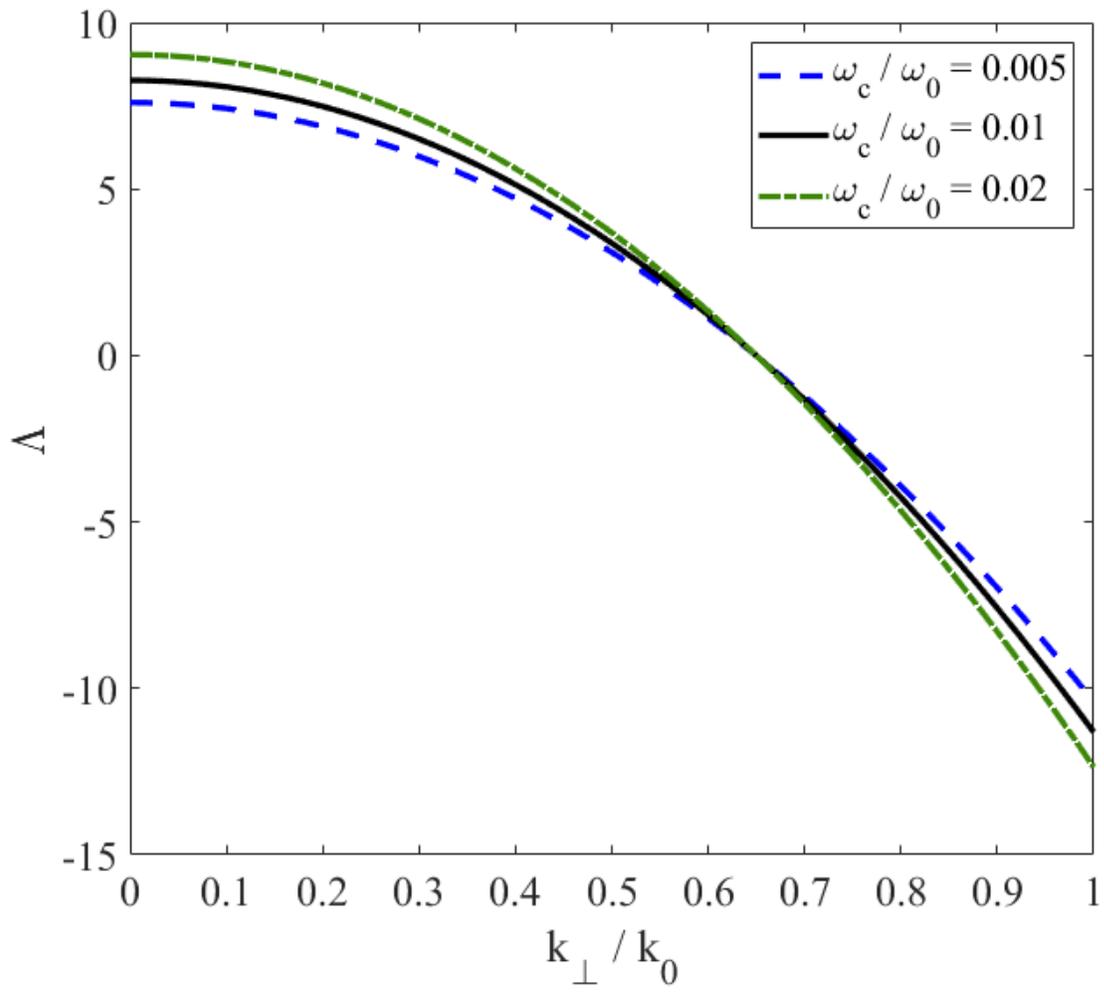

**Fig. 3**

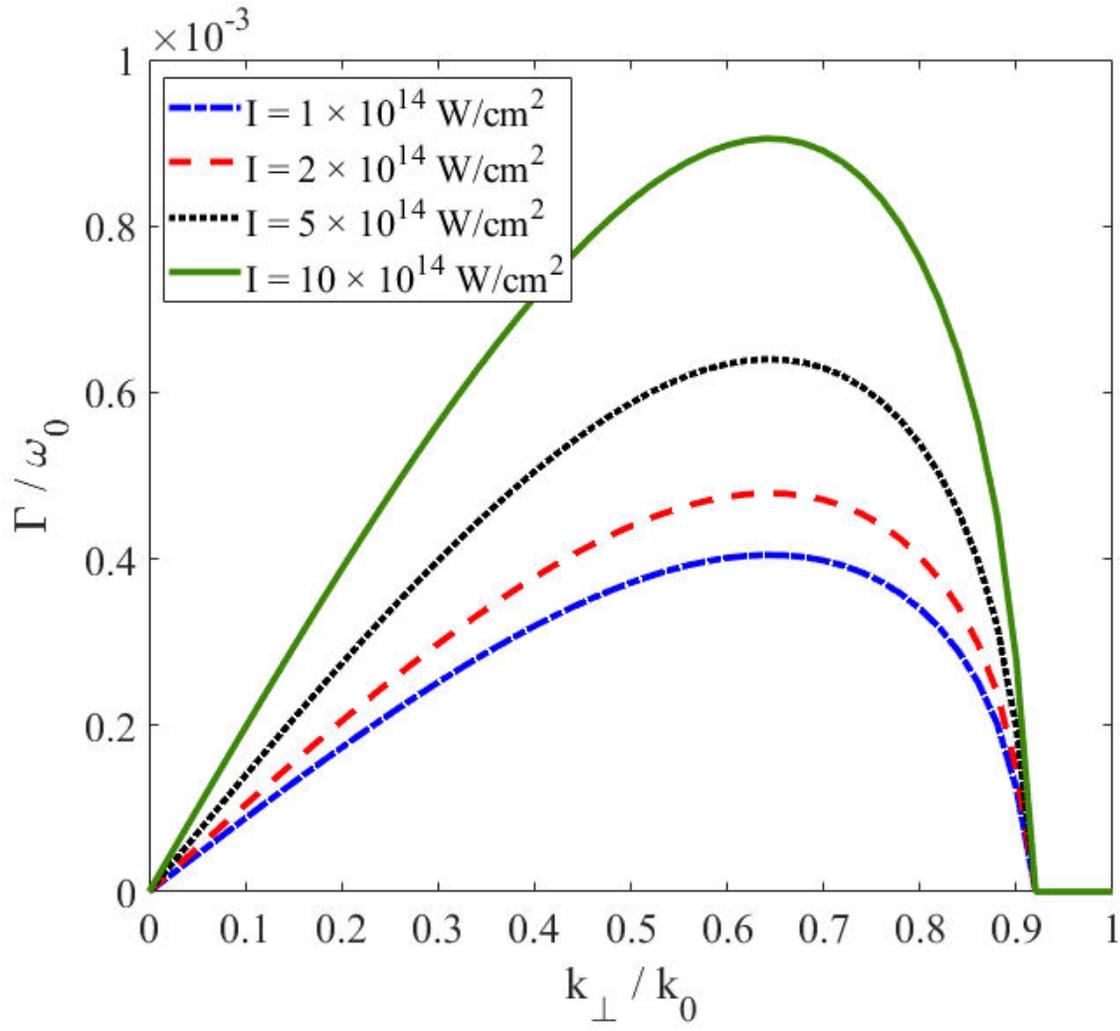

**Fig. 4**

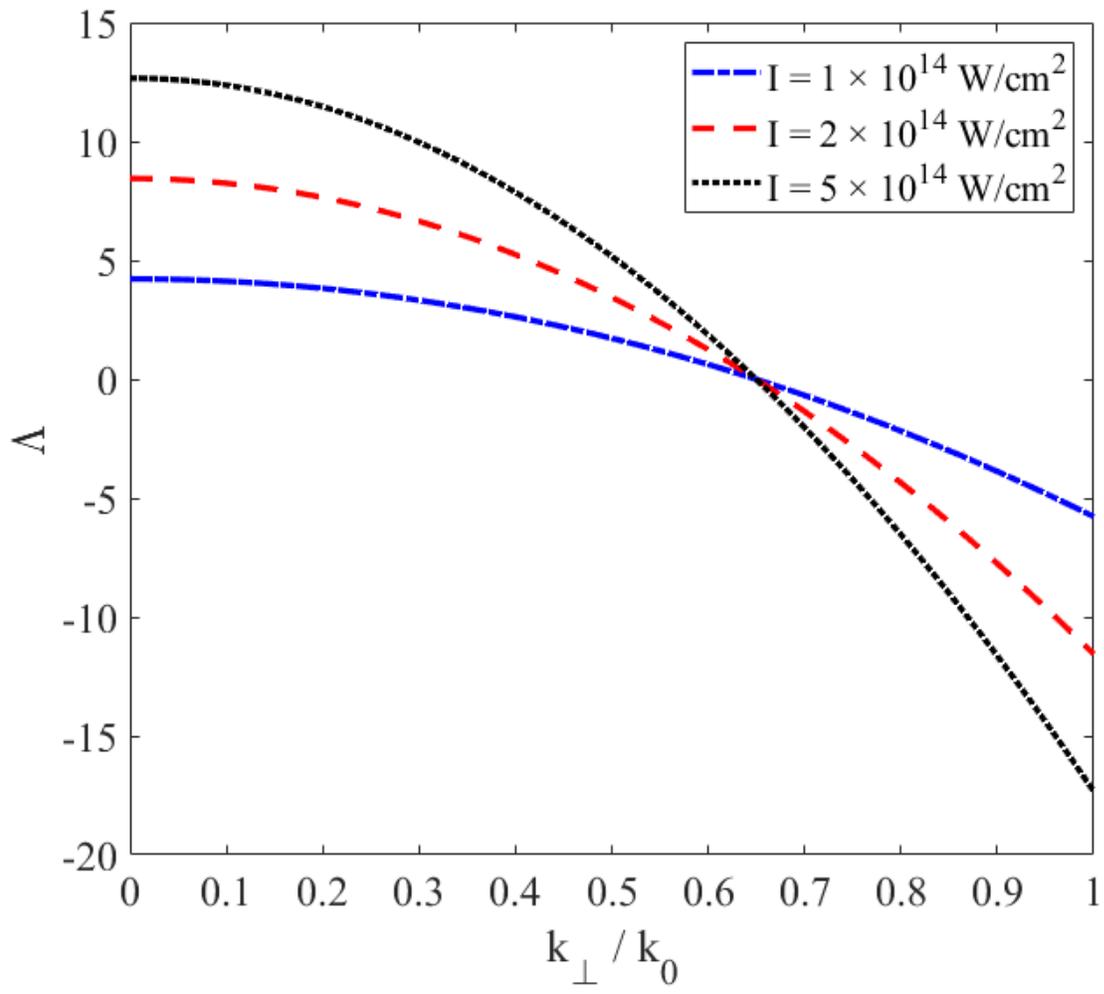

**Fig. 5**

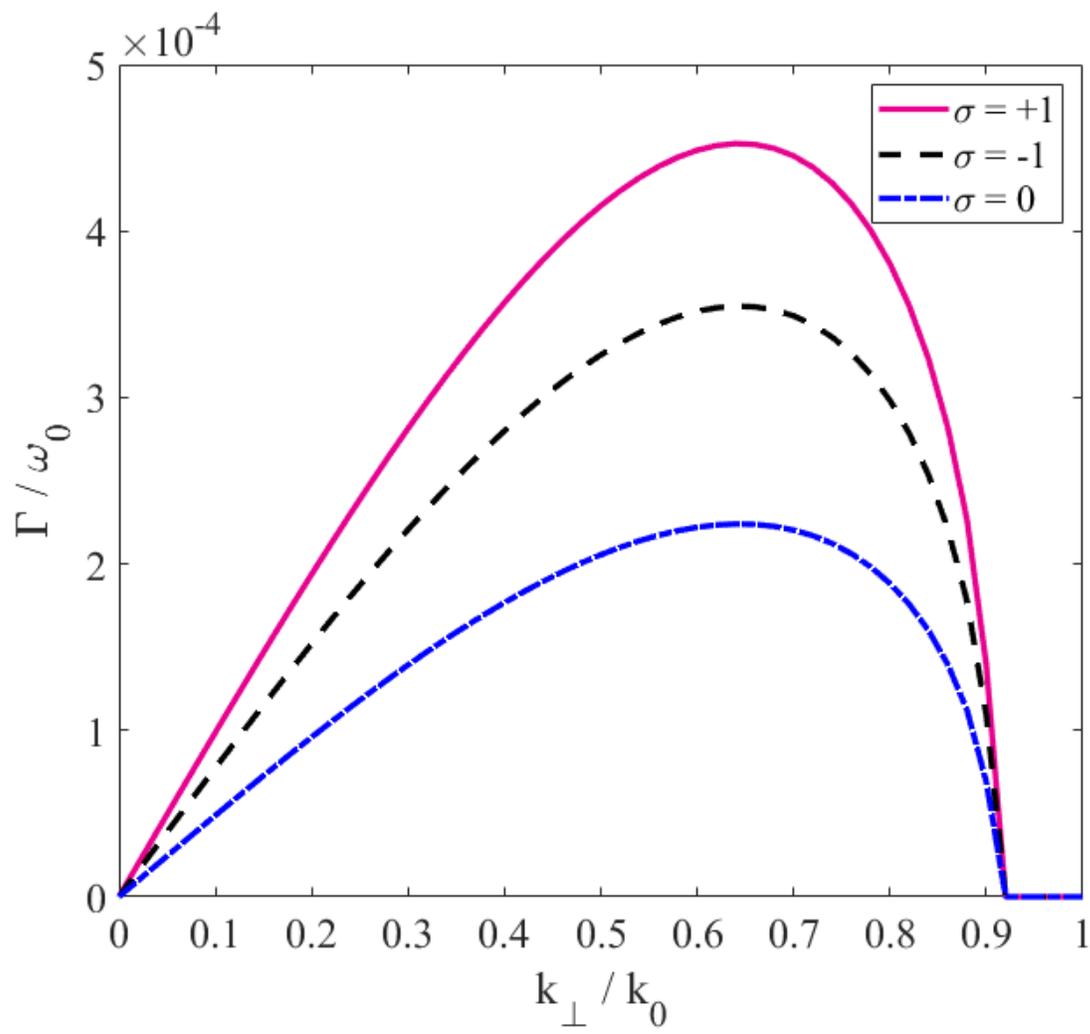

**Fig. 6**

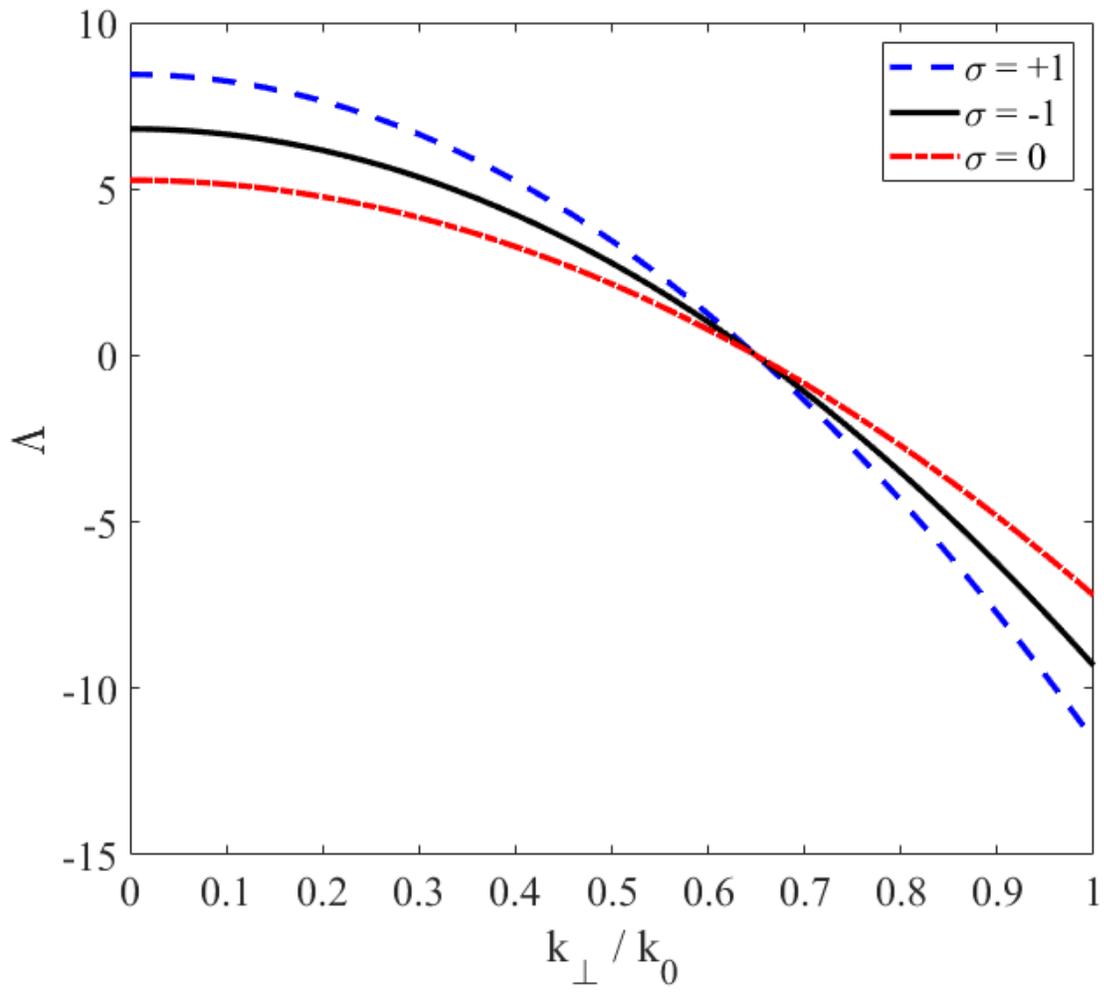

**Fig. 7**

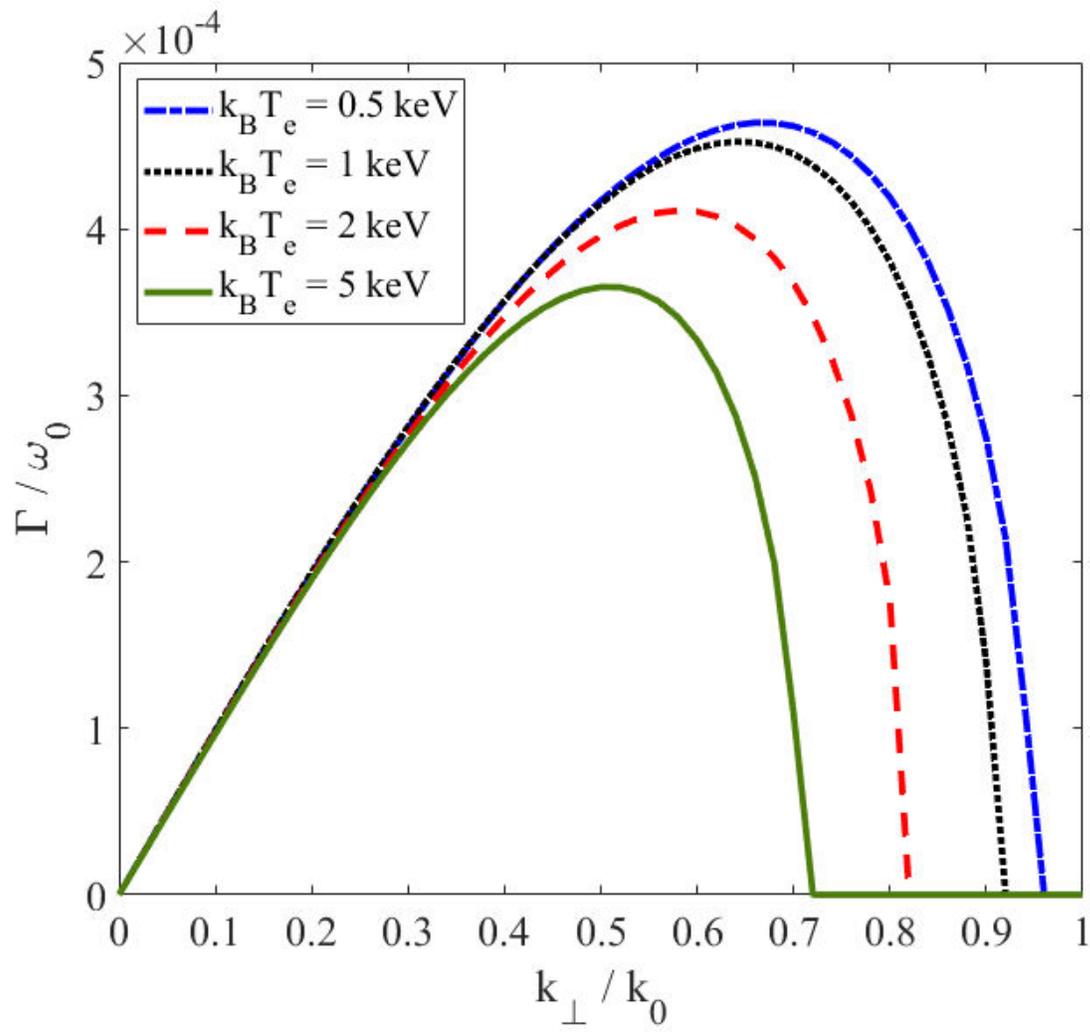

**Fig. 8**

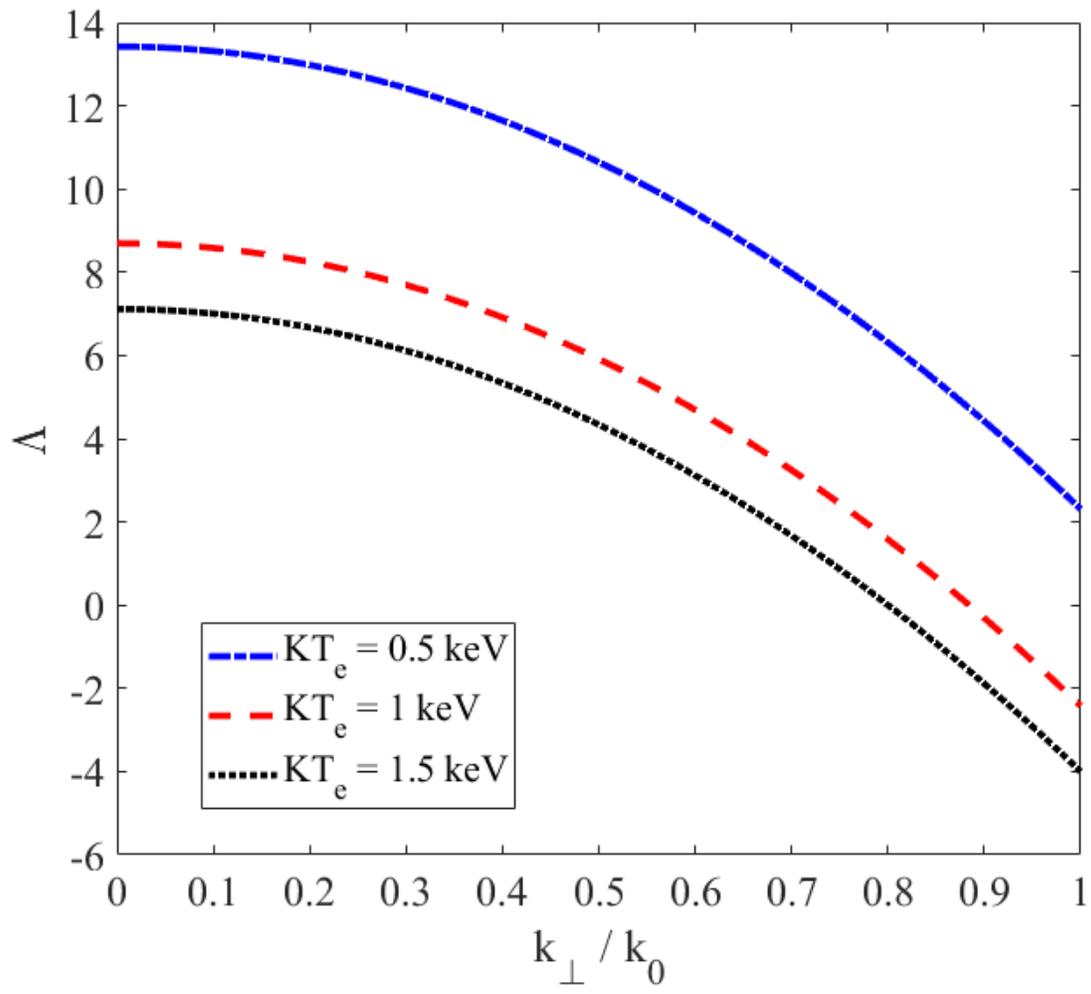

**Fig. 9**

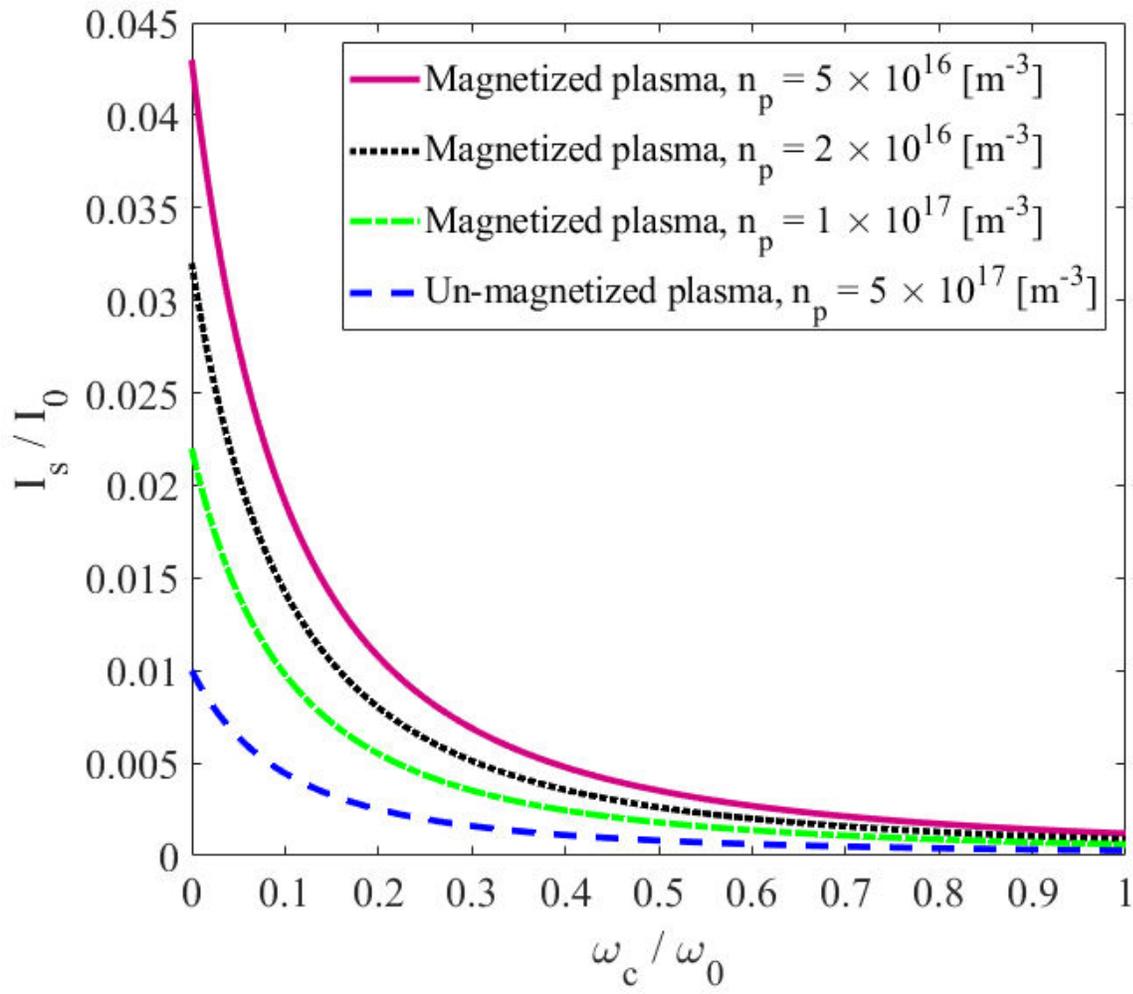

**Fig. 10**